# Improving 3D Synthetic Jet Modeling in a Crossflow

**Haonan Howard Ho[1]**

University of Toronto

King's College Road, Toronto, Ontario M5S 3G8, Canada

howard.ho@mail.utoronto.ca

**Ebenezer Ekow Essel**

Concordia University

1515 St. Catherine W. Montreal, Quebec H3G 1M8, Canada

ebenezer.essel@concordia.ca

**Pierre Edward Sullivan**

University of Toronto

King's College Road, Toronto, Ontario M5S 3G8, Canada

pierre.sullivan@utoronto.ca

ASME Fellow

---

[1] Corresponding author.

**ABSTRACT**

*Three different circular synthetic jet modeling inlet conditions are studied for a turbulent crossflow. The study examines the differences when modeling the whole SJA, neck-only or jet-slot-only under constant actuation frequency (f = 300 Hz) and crossflow blowing ratio ($C_B$ = 0.67). Phase-averaged and time-averaged results reveal that both whole SJA and neck-only methods generated nearly identical flow fields. For the neck-only case, a notable reduction in computational cost is achieved through the implementation of an analytical jet profile. The jet-slot-only method, on the other hand, introduces reversed flow during the ingestion cycle, leading to the injection of false-momentum into the crossflow. However, the false-momentum primarily affects the flow immediately downstream of the jet exit, with the boundary layer profile recovering rapidly. A parametric study highlights the importance of maintaining a volume ratio less than 1 of ingested to modeled neck volume to prevent the creation of false-momentum.*

## 1  INTRODUCTION

Synthetic jet actuators (SJAs) are zero-net-mass-flux (ZNMF) devices with wide applications, including flow separation control[1–4], combustion chamber fuel mixing[5,6], jet vectoring[7], cooling of electronic devices and heat transfer enhancement[8–11]. The key advantage of SJA is their ability to operate without a dedicated fluid supply. Synthetic jet in a crossflow is also of fundamental importance in understanding the three-dimensional (3D) unsteady interactions between vortex rings and boundary layers or separation bubbles. Accordingly, significant research has been performed to characterize and optimize SJAs using both experimental and numerical techniques[12–14]. However, for the numerical approaches, the inherent challenge has been the modeling of the periodic expulsion of the synthetic vortex rings to agree with the experiments. The goal of the present study is to compare various SJA modeling strategies to better understand their impact on the jet-crossflow interactions and computational resources.

A schematic of an SJA in a crossflow boundary layer is shown in **Fig. 1**. The crossflow has a freestream velocity of $U_\infty$ and a boundary layer thickness of $\delta$. The SJA consists of a cavity where periodic vibrations of the diaphragm with amplitude $a$ ingests and expels part of the crossflow, effectively transferring momentum to the boundary layer in the form of periodic vortex rings. The vortex rings evolve and diffuse away from the jet exit through interactions with the crossflow.

Particle image velocimetry (PIV), hot-wire anemometry and dye visualization have been used to investigate the flow characteristics of synthetic jets in both quiescent[15–17] and crossflow[18–21] conditions. Important parameters used to define the flow characteristics includes jet blowing ratio ($C_B = \overline{U}_J/U_\infty$), stroke length ($L = \frac{\overline{U_J}}{fd}$) and momentum coefficient ($C_\mu = \frac{\rho_j \overline{U_J}^2}{\rho_\infty U_\infty^2} \frac{d}{\theta_0}$), where $\overline{U}_J$ is the average jet velocity during expulsion cycle, $f$ is the actuation frequency, $d$ is the jet diameter, $\rho_j$ is the jet fluid density, $\rho_\infty$ is the crossflow boundary layer fluid density and $\theta_0$ is the momentum thickness of the unactuated boundary layer. For a low momentum jet ($C_B < 0.35$), hairpin vortices are found to remain close to the wall, but as the $C_B$ increases, vortex rings are formed instead, which undergo tilting and penetrate rapidly into the boundary layer[22,23].

Numerical simulations based on Unsteady Reynolds-Averaged Navier-stokes (URANS)[24,25] and Large Eddie Simulations (LES)[26,27] have been used to augment planar (2D) experiments and provide additional insight on the 3D evolution of the vortex rings and the near-wall region. The approaches for modeling the SJA can be classified into three main categories: modeling of (i) whole SJA, (ii) neck-only and (iii) jet-slot-only (**Fig. 2**). For the whole SJA method, a dynamic mesh is used to model the diaphragm displacement[28,29] defined as

$$s = a(1 - (r/r_o)^2)sin(2\pi f t) \qquad (1)$$

where $a$ is the maximum vibration amplitude, $r$ is the local radius, $r_o$ is the diaphragm radius, $f$ is the actuation frequency and $t$ is time. Pasa et al.[28] used the dynamic mesh method to conduct 2D URANS simulations of an array of four SJAs in a quiescent flow. They found that phase optimized SJAs significantly increased the jet velocity at far field ($y/d$ = 20). Ho et al.[23] recently conducted simulations on a circular SJA issuing into a turbulent boundary layer, the dynamic mesh method was used to investigate the effect of jet momentum under fixed crossflow characteristics. The whole SJA dynamic mesh method is widely used, however, the computational cost associated with grid displacement has prompted the exploration of alternative cost-effective SJA modeling techniques[30–32].

The neck-only method is an alternate approach that bypasses the need for modeling the cavity and motion of the diaphragm of the SJA. Here, a temporal- and spatial-varying velocity field is applied to the bottom of the neck of the SJA[33–35]. Palumbo et al.[30] performed DNS simulations using the neck-only method to investigate SJAs as vortex generators for turbulence transition. The analytical Womersley solution for a pulsating laminar pipe flow was specified as velocity input for the neck:

$$v(r,t) = V_j \cdot Real\left\{\left[1 - \frac{J_o(i^{3/2}W_o r)}{J_o(i^{3/2}W_o)}\right]e^{i\omega t}\right\} \qquad (2)$$

where $V_j$.is the maximum jet center-line velocity, $\omega$ is the actuation frequency in radians, $W_o$ is the Womersley number ($W_o = d\sqrt{\omega/4v}$), and $J_o$ is the zeroth order Bessel function. They found that the jet characteristics is independent of neck height ratio for $h/d \geq 5$, where $h$ is the height of the neck. LES simulations using neck-only method were conducted by Asgari and Tadjfar[36] to study phase-difference effect on adjacent rectangular SJAs on a rounded ramp. A temporal varying top-hat like jet profile was assigned at the bottom of the neck. Validation of the SJA in a quiescent flow showed that the neck-only approach produced result that agree with DNS based on whole SJA method[37] and experiments[38].

For the jet-slot-only method, both the cavity and the neck of the SJA are omitted[4,26,39,40]. A temporal and spatial varying velocity profile is applied to the jet exit on the control surface, further reducing the computational cost and geometric restriction. Bai et al.[32] used the jet-slot-only method and LES to study flow control around a square cylinder, while Wang et al.[31] used the approach to develop a deep reinforcement learning closed-loop control scheme for separation control on a 2D NACA 0012 airfoil. The closed-loop scheme developed[31] eliminated vortex shedding and increased the lift to drag coefficient.

Raju et al.[41] conducted a prior investigation into the three different modeling techniques, with a particular emphasis on 2D SJAs in attached grazing flow and canonical separated crossflow. Their work compared the whole SJA method with a radius based jet profile *(1)*, both uniform ($v = f(t)$) and sink like jet ($u, v = f(x,t)$) to a neck-only model, and a uniform jet on the jet-slot-only model. It was concluded that the sink like jet profile on a neck-only model had the best agreement when compared to the whole SJA method, while reducing the computational cost significantly. The jet-slot-only model was able to predict the correct trend in the separation control, however, it underpredicted the separation bubble and the deviation increased rapidly as the jet Reynolds number was increased.

While the three techniques have been used extensively and briefly explored before, our understanding of their relative strengths and weaknesses in circular SJA is insufficient, due to the limited comparative studies that examine these approaches in 3D. Therefore, the objective of this study is to evaluate the effectiveness of a circular SJA with the whole SJA, neck-only and jet-exit only modeling approaches in a turbulent crossflow using 3D URANS. The crossflow Reynolds number ($Re_{\theta_0} = 900$), blowing ratio ($C_B = 0.67$) and actuation frequency ($f = 300$ Hz) were kept constant for all test cases. The goal is to provide a general guide for modeling SJA with acceptable compromise between accuracy and computational cost.

## 2    NUMERICAL METHODS

Following previous study[23], the 3D URANS simulations were performed using the Launder-Sharma Low-Reynolds number $k-\varepsilon$ model [42] in OpenFOAM V1912[43]. The PisoFOAM solver was used to solve the discretized governing equations based on finite volume method. To ensure the stability, a first-order accurate upwind scheme was used for the turbulence quantities while a second-order scheme was used for the convection terms. The convergence criterion for all variables was set to $10^{-6}$. The first-order Euler time marching scheme was used for temporal discretization and the time step $\Delta t$ was controlled automatically to ensure the Courant number is less than unity during computation.

### 2.1    Model Setup, Boundary and Test Conditions

Three different methods for modeling SJA are investigated, **Fig. 2**. For the whole SJA method, the dimensions of the SJA is based on the experiments by Feero et al.[44], where $d$ = 2 mm and the neck height, $h/d$ = 5. The cavity has a diameter of 30.8 mm and a height of 10 mm. A dynamic mesh function was applied to model the diaphragm displacement *(1)* . For the neck-only approach, the SJA cavity computational domain is omitted and the analytical Womersley solution *(2)* is applied to the bottom of the neck[30]. For the jet-slot-only approach, the Womersley solution is applied directly at the jet exit, both SJA cavity and neck are omitted.

The computational domain of the whole SJA in crossflow is presented in **Fig. 3**. The SJA jet exit center ($x/d$ = 0, $y/d$ = 0) was located at the mid-span of the duct and 25$d$ downstream of the inlet. A hybrid computational grid was generated for the duct section of the crossflow domain, inflation layers were assigned to both top and bottom walls. The cells around the SJA jet exit and immediate downstream were highly refined, coarsen cells were instead assigned to the free stream portion of the duct. Within the SJA, a butterfly structured grid is constructed for both the neck and the cavity. In all cases, the duct domain has a length of 200$d$, width of 20$d$ and height of 38$d$. A velocity inlet and pressure outlet boundary condition (BC) were applied to the duct. All walls were assigned no-slip condition while the sides of the domain were assigned symmetry. A separate duct flow simulation was performed to obtain the inlet velocity profile of the turbulent boundary layer.

A mesh independence study[45] was performed on both whole SJA and neck-only case where three different grids with increasing cell counts were tested. The mesh was refined in all three directions in each step, especially in regions of high shear layer interaction and within the inflation layers. **Table 1** shows the mesh independence study with the maximum center-line jet velocity $U_{cl}$ and the time-averaged displacement thickness $\delta^*$ at $x/d$ = 10 as key parameters. The sampling began after 10 cycles to accommodate flow development. The time averaged displacement thickness was obtained at a rate of 32 samples per actuation cycle over 2 cycles, then the flow field was time averaged. Grid II was selected for this work as the accuracy was acceptable and the computational cost was moderate. The inflation layers of the medium density mesh consisted of 40 layers, a first cell heigh of 75μm and a growth rate of 1.07. The maximum y⁺ of the walls from the medium mesh was roughly 1.

Validations of the turbulent boundary layer and the whole SJA approach in quiescent with experimental data were previously conducted[23]. The boundary layer momentum thickness Reynolds number was $\text{Re}_{\theta_0} = 900$ and had a freestream velocity $U_\infty$ of 7.24 m/s at *x/d* = -5. The SJA operates at fixed actuation frequency of 300 Hz and momentum coefficient $C_\mu = 0.46$. The diaphragm vibration amplitude for the whole SJA case was 34 $\mu m$, leading to an averaged jet velocity $\overline{U_J}$ of 4.8 m/s, a jet Reynolds number $\text{Re}_{\overline{U_J}}$ of 642 and a maximum center-line jet velocity $U_{cl}$ of 7.9 m/s. In the neck-only case, the maximum center-line velocity assigned for the Womersley solution $V_j$ based BC was adjusted to accommodate flow developing within the neck. To match the flow characteristics of the SJA, a value of $V_j = 8.5$ m/s was applied to the BC for the neck-only case, which evolved into a maximum center-line velocity of $U_{cl} = 7.9$ m/s at *y/d* = 0.075. No jet matching preliminary simulation was performed for the jet-slot-only case, the Womersley solution with $V_j = 7.9$ m/s was applied directly to the jet exit. The computational cost for one SJA actuation cycle in the crossflow for the neck-only case was reduced by 72% compared to the dynamic mesh whole SJA case, similar to those reported by Raju et al.[41]. The jet-slot-only case also measured a significant cost reduction of 70%, which was slightly less than the neck-only case and will be explored in the results section.

Sampling of the crossflow simulations begin after 2 flow-through-cycles ($l/U_\infty$, where $l$ is the length of the computational domain of the crossflow). Time-averaged results consist of 160 samples at a rate of 32 per actuation cycle, phase-averaged results are taken from averaging of same phase data over five actuation cycles.

## 2.2 Validations of SJA BC

**Fig. 4** displays the center-line jet velocity at *y/d* = 0.075 over one actuation cycle, where $\theta$ is the phase angle respective to the actuation signal. The neck-only case results are shifted linearly by 10 degrees to match the maximum center-line velocity of the whole SJA case. This phase difference is expected as the neck-only case BC was applied closer to the sampling location.

The radial phase-averaged velocity jet profiles during peak expulsion ($\phi = 90°$) and ingestion ($\phi = 270°$) phase of the neck-only case is presented in **Fig. 5**, compared with the whole SJA dynamic mesh results and experimental data in quiescent flow[15]. The jet profile of the neck-only case from the Womersley solution is in great agreement with the whole SJA approach, similar to the center-line velocity plot in **Fig. 4**.

## 3   RESULTS AND DISCUSSION

### 3.1   Instantaneous Flow Structure

Q-criterion contours are often used to analyze 3D unsteady vortical structures, where Q represents the second invariant of the velocity gradients tensor[46]. **Fig. 6** displays the instantaneous iso-surface of Q-criterion at $\phi = 0°, 90°, 180°$ and $270°$ for all three cases, colored by normalized spanwise vorticity. From all three approaches, a downstream tilting jet column is formed during expulsion phase ($\phi = 90°$ and $180°$). On the upstream side of the jet columns, vortex loops are formed due to shear layer interactions between the jet column and the boundary layer[23][47]. As the SJAs alternate into ingestion cycle ($\phi = 270°$), the vortex loops have decayed entirely, leading to the formation of a tilted vortex ring (VR) attached by a trailing vortex pair[18]. For the whole SJA and neck-only cases, the trailing vortex pair stretches as the primary VR convects downstream ($\phi = 0°, 90°$ and $180°$), resembling the two leg-like structure of a hairpin vortex[48,49]. The trailing vortex pair of the jet-slot-only case evolves into a single leg-leg like structure along the symmetry plane, leading from the SJA slot center during the ingestion cycle ($\phi = 0°$ and $270°$). Compared to the trailing vortex pair of the whole SJA and neck-only case, the single leg-like trailing vortex of the jet-slot-only case decays much quicker and is no longer visible by $\phi = 270°$, only the primary VR remains.

Close to the bottom wall, a horseshoe vortex is observed on the upstream side of the jet exit across all three cases during expulsion cycle ($\phi = 90°$ and $180°$). During the peak ingestion phase ($\phi = 270°$), a near-wall tertiary vortex along the symmetry plane is formed downstream of the jet exit. The whole SJA and neck-only cases near-wall tertiary vortex remained consistent as it convects downstream, while the jet-slot-only case near-wall tertiary vortex appeared less structured. However, all three approaches eventually resulted in a tilted VR by $\phi = 270°$, in agreement with the experimental work of Jabbal and Zhong[22] at similar crossflow conditions.

### 3.2    Phase-averaged Symmetry Plane Flow Profile

**Fig. 7** displays phase-averaged symmetry plane streamwise velocity contours normalized by free-stream velocity across the three cases. Streamline plots are superimposed on the velocity contours. During the peak expulsion phase ($\phi = 90°$), a reversed flow region on the downstream side of the jet exit is observed, the area of the reversed flow region from different cases are similar in size but not identical, which is contributed by a slight phase difference due to the distance between the jet exit and the plane which actuation is applied. The reversed flow region area reaches maximum size during the switch over phase ($\phi = 180°$) as ingestion cycle begins[23]. During the ingestion phases ($\phi = 270°$ and $0°$), a physically non-accurate expelling jet is observed from the jet-slot-only case, originated from jet exit, and continues to add false-momentum into the boundary layer, in contrast to the whole SJA and neck-only case. The streamline plots of the jet-slot-only case are also in disagreement with the other two cases, where the resultant flow of the jet-slot-only case appeared to be more uniform along the boundary layer. This false-momentum of the jet-slot-only case inaccurately increased the near-wall momentum of the boundary layer, which can lead to overprediction of the effectiveness of an SJA. The false-momentum observed during the jet-slot-only ingestion cycle also contributed to additional computational cost when compared to the neck-only case, as the time-step of the solve was lowered to maintain stability.

As the expelled structure evolves downstream and passes $x/d$ = 5, all three cases presented a region of momentum deficit at a distance intermediate to the bottom wall ($y/d$ < 4), contrasted by an increase of streamwise flow velocity at $y/d$ = 5. This is the result of the primary VR entraining flow from the far field into the boundary layer, which has been suggested to be responsible for suppressing shear layer roll up of airfoils[50,51].

### 3.3    Time-averaged Flow Profile

To examine the impact on the boundary layer in the three different cases, time averaged streamwise velocity and turbulent kinetic energy $k$ profiles are generated along the symmetry plane, **Fig. 8**. The whole SJA and neck-only case have near identical time-averaged profiles in both streamwise velocity and $k$ at all locations examined. For the Jet-slot-only case, the physically inaccurate expelled jet formed during the ingestion cycle added significant near-wall momentum close to the jet exit ($x/d$ = 1). The increase of momentum is still noticeable at $x/d$ = 5 and had diffused further into the boundary layer ($y/d$ = 3), but the local disagreement is less severe compared to $x/d$ = 1. By $x/d$ = 10, the added momentum of the jet-slot-only case had largely diffused and resulted in a boundary layer profile in good agreement with the other two cases.

For the time-averaged $k$ profile immediately downstream of the SJA ($x/d$ = 1), the peak rests close to the bottom wall at $y/d$ = 0.5, followed by a second peak further into the boundary layer ($y/d$ ~ 2). While the location of the primary $k$ peak of the jet-slot-only case generated was the same as the other two cases, the predicted maximum $k$ was 60% greater. The increased $k$ of the jet-slot-only case however was only observed near the jet exit, by $x/d$ = 5 the $k$ profile had minor regional deficit compared to the other two cases, similar behavior was observed further downstream ($x/d$ = 10).

Time averaged streamwise velocity is presented for the $x$-$z$ plane at $y/d$ = 0.5 to examine the spanwise influence of the different cases (**Fig. 9**). Similar to the symmetry plane results, the whole SJA and neck-only cases generated near identical time averaged flow fields. For the jet-slot-only case, a larger time-averaged streamwise velocity is observed immediate downstream of the jet exit. Comparison between the jet-slot-only and whole SJA case in **Fig. 10** showed that the near wall velocity was increased by more than 50% between 1 < $x/d$ < 3. Despite the significant increase of near wall momentum of the jet-slot-only cases, the disparity is primarily along the symmetry plane and dissipates rapidly, returning to within 15% of the whole SJA results by $x/d$ = 5.

Streamwise velocity distribution along the span further away from the bottom wall at $y/d$ = 1.5 is shown in **Fig. 11**. The spanwise effect of the three cases are mostly confined between -3 < $z/d$ < 3, a region of streamwise velocity deficit is observed along the symmetry plane (-1 < $z/d$ < 1). The whole SJA and neck-only cases both predicted a symmetric profile, while the jet-slot-only case predicted a non-symmetric profile at $x/d$ = 5. The false-momentum also resulted in the jet-slot-only case having greater momentum near the symmetry plane. The non-symmetric behavior of the jet-slot-only case was no longer present by $x/d$ = 10, and the symmetry plane momentum was only slightly greater than the other two cases (~5%).

The skin friction coefficient is often used to explore potential separation control of SJAs, **Fig. 12** compares the time-averaged skin friction coefficient of the three cases. On the upstream side of the jet exit for the whole SJA and neck-only cases, a horse-shoe shaped profile of increased skin friction is observed. Area immediate downstream of the jet exit along the symmetry plane has a lower skin friction coefficient till $x/d$ = 2, followed by an increase of skin friction again. The two trailing vortex pair observed has also resulted in greater spanwise coverage (-1< $z/d$ <1), displaying similarity with a hairpin vortex[22,23]. A different behavior was observed for the jet-slot-only case downstream of the SJA, the false-momentum had increased the skin friction along the symmetry plane significantly, resembling the profile of a high momentum VR instead[22,23].

### 3.4    The Volume Ratio

The comparison of the previous sections displayed that when the neck of a circular SJA is not modeled, physically inaccurate behavior occurs at the jet exit during ingestion cycle, which injects false-momentum into the crossflow. This false-momentum near the jet exit has significant effects on near wall variables such as wall shear stress, but the effect overall dissipates rapidly. To identify a key variable to prevent unrealistic behavior when modeling SJAs, a parametric study of the neck-only case in crossflow with increasing neck height is investigated, $h/d$ = 0.5, 2, 4 and 6. Prior to the parametric study, preliminary simulations in quiescent using the various neck height ratios were conducted to obtain a consistent maximum center-line velocity $U_{cl}$ for all test cases. All sampling rate and procedure conducted in this section is consistent with that mentioned in **Section 2.1**.

The instantaneous streamwise and transverse velocity is sampled at the jet exit center ($x/d$ = 0, $y/d$ = 0.075) and plotted for all four neck heights relative to phase angle $\Phi$ in **Fig. 13**. The streamwise velocity plot **Fig. 13**(a) displayed that mild unsteadiness was observed during part of the expulsion cycle ($\Phi = 0° − 135°$), however, the 4$d$ and 6$d$ cases show near identical results while slight deviation is observed for the 2$d$ case. The shortest neck case of 0.5$d$ shows significantly different streamwise velocity during the transition from ingestion to expulsion ($\Phi = 300° − 45°$), the unsteadiness occurred earlier and much more vigorous than the other cases. A slight phase shift of the shortest neck case (0.5$d$) is expected as the Womersley solution is applied significantly closer to the sampling location, but a change of peak values and overall trend is not anticipated. For the transverse velocity **Fig. 13**(b), the 4$d$ and 6$d$ cases again displayed consistent behavior, slight disagreement is present from the 2$d$ case, while the 0.5$d$ case has a profile that closely resembles the quiescent center-line-line velocity profile (**Fig. 4**). The quiescent flow like transverse velocity profile of the 0.5$d$ case indicates the Womersley solution BC was applied too close to the crossflow, where insufficient modeling domain was provided to accommodate the crossflow interactions.

Radial jet profiles are displayed in **Fig. 14** for all neck heights at peak ingestion ($\Phi = 0°$), expulsion ($\Phi = 270°$) and the transition phases ($\Phi = 0°, 180°$). The deviation of the 0.5$d$ case is not confined to the jet center-line only, as seen during the initiation of expulsion phase ($\Phi = 0°$) in both streamwise and transverse velocity jet profile. Similar to the center-line velocity, a general convergence behavior with increasing neck height is observed at $\Phi = 0°$. However, the ingestion cycle profiles ($\Phi = 180° − 270°$) are in good agreement even for the shortest neck 0.5$d$ case. The reasonable agreement of the shortest neck case during ingestion cycle provided a contrast to the jet-slot-only case investigated explored in **Section 3.2**, the false-momentum was still added in the shortest neck 0.5$d$ c case, but it was less severe and required additional time to convect to the crossflow region.

To examine the neck volume's impact on the convergence behavior, rather than solely on the neck height, the parametric study also includes investigation on the modeled neck volume effect of the neck-only cases. The sum of flow ingested into the SJA is first sampled from the whole SJA case, a plane located mid-neck ($y/d$ = -2.5) was chosen to minimize non-axial flow within the neck, then divided by the neck volume modeled of the parametric study cases as presented in **Table 2**. A greater than unity volume ratio means that the ingested volume of fluid into the SJA is greater than that of the modeled neck volume. It was found that the volume ratio for the 4$d$ and 6$d$ cases are both less than 1, while the 2$d$ and 0.5$d$ cases had a volume ratio of 1.9 and 7.6, respectively. It is, therefore, hypothesized that to avoid unrealistic near jet behavior, a volume ratio of less than unity is necessary.

To validate the volume ratio dependence on neck volume rather than neck height, two cases with different jet diameter based on the neck-only method are simulated. The two different neck diameter SJA shares the same actuation frequency ($f = 300\ Hz$). In order to ensure the fundamental crossflow interactions of the two different diameter SJAs are comparable, the peak neck momentum was used instead jet velocity or jet flow rate. Preliminary simulations in quiescent flow ensures both SJA has a peak neck momentum within 5% of the whole SJA case (sampled at *y* = -5 mm). The narrow case has a neck diameter *d* of 1.4 mm while the wide case has a neck diameter *d* of 2.8 mm, the maximum jet center-line velocities $V_j$ applied for the Womersley solution BCs are 12.3 m/s for the narrow case, and 5.9 m/s wide case. Both narrow and wide cases share the same non-dimensional neck height (*h/d* = 4), while the volume ratios are assigned 2.0 for narrow and 0.5 for the wide neck case. Prior results have shown that the false-momentum dissipates rapidly in a crossflow, therefore the focus on the narrow and wide cases are placed within the neck region immediate to the Womersley solution boundary.

Phase averaged transverse velocity $U_y$ and streamwise velocity $U_x$ are sampled 100 μ*m* from the bottom of the neck and displayed in **Fig. 15** for both narrow and wide cases. The narrow case's transverse velocity $U_y$ at peak ingestion phase ($\Phi = 270°$) and initiation of expulsion phase ($\Phi = 0°$) is not axisymmetric, **Fig. 15**(a), while the wide case's profile is axis-symmetric. The Womersley solution prescribes an axial symmetric analytical profile, the non-symmetric profile of the narrow case immediate from the boundary suggests failed implementation of the BC. Furthermore, the narrow case also displayed significant streamwise velocity $U_x$ during ingestion cycle **Fig. 15**(b), which is perpendicular to the axial flow direction assigned by the BC and indicates formation of a vortex at the bottom of the neck, resulting in reversed flow.

The impact of the insufficient volume ratio narrow case also extends to turbulence quantities (**Fig. 16**). The Womersley solution profile describes a laminar pulsing pipe flow and does not produce significant turbulence, however, the narrow case predicted high amount of $k$ and turbulence dissipation rate $\varepsilon$ at the bottom of the neck during peak ingestion phase ($\Phi = 270°$). As both the wide and narrow neck cases shared the same normalized neck height, the results confirm that the neck volume is the key modeling parameter, rather than the neck height.

## 4    CONCLUSION

In this work, three different methods for modeling 3D circular SJA in a crossflow boundary layer are investigated using Unsteady Reynolds-Averaged Navier-Stokes and Low-Reynolds number turbulence model. The first, the whole SJA method models the entire SJA computational domain, a dynamic mesh function is used to model the diaphragm displacement. The second, the neck-only method models only the neck portion of the SJA, an analytical profile is assigned to the bottom of the neck. The third, the jet-slot-only method applies the analytical profile directly on the jet exit, omitting the SJA neck and cavity.

All three methods predicted the formation of a tilted vortex ring and the horseshoe vortex. The whole SJA and neck-only methods predicted a trailing vortex pair attached to the tilted vortex ring, a single leg-like structure is observed for the jet-slot-only case instead and it dissipates rapidly. The difference in the structure of the jet-slot-only case is found to be the result of reversed flow at the at the bottom of the neck during ingestion cycle, where significant false-momentum is added to the boundary layer. The time-averaged boundary layer profile showed that both the whole SJA and neck-only cases exhibit identical profiles in both the symmetry and spanwise planes. The false-momentum from the jet-slot-only case increased the near-wall ($y/d < 1$) momentum immediately downstream of the jet exit. However, the added false-momentum dissipates rapidly, and the results are consistent with the whole SJA and neck-only cases by $x/d$ = 10. Time-averaged skin friction coefficient from the whole SJA and neck-only cases were in good agreement, but the jet-slot-only case overestimated the skin friction coefficient immediately downstream of the jet exit along the symmetry plane.

A neck volume parametric study was conducted on the neck-only approach with neck height ratios ($h/d$ = 0.5 - 6). A convergence behavior was observed with increasing neck volume. A critical volume ratio is, therefore, proposed as the ingested flow volume should be less than the modeled neck volume to avoid physically inaccurate SJA behavior near the jet exit. The volume ratio was tested with two different neck diameters with the neck-only SJA method in crossflow. The results showed that the case with insufficient neck volume (volume ratio = 2) had resulted in non-symmetric flow profile and reversed flow at the bottom of the neck during ingestion cycle. Meanwhile, the sufficient neck volume case (volume ratio = 0.5) resulted in axis-symmetric flow and no physically inaccurate behavior at the same region of the neck. The results of the different neck diameter cases indicate the neck volume is more important than the neck height in partial modeling of circular SJA in crossflow.

**ACKNOWLEDGMENT**

The authors acknowledge the support from the Natural Sciences and Engineering Research Council of Canada and CMC Microsystems. Computations were performed at the SciNet HPC Consortium.

**NOMENCLATURE**

| | |
|---|---|
| $a$ | Diaphragm vibration amplitude, mm |
| $C_B$ | Blowing ratio; $\overline{U}_J / U_\infty$ |
| $C_\mu$ | Momentum coefficient; $\rho_J \overline{U}_J^{\,2} D / \rho_\infty U_\infty^2 \theta_0$ |
| $d$ | Jet and neck diameter, mm |
| $f$ | Synthetic jet actuation frequency, Hz |
| $f_o$ | Vortex shedding frequency, Hz |
| $H$ | Cavity height, mm |
| $h$ | Neck height, mm |
| $J_o$ | Zeroth order Bessel function |
| $k$ | Turbulence kinetic energy, m²/s² |
| $L$ | Stroke length; $\overline{U}_J / f D$ |
| $l$ | Duct computational domain length, mm |
| $\mathrm{Re}_c$ | Chord Reynolds number; $U_\infty c / \nu$ |
| $\mathrm{Re}_{\overline{U}_J}$ | Synthetic jet Reynolds number; $\overline{U}_J d / \nu$ |
| $\mathrm{Re}_{\theta_0}$ | Momentum thickness Reynolds number oof unactuated boundary layer; $U_\infty \theta_0 / \nu$ |
| $r$ | Local radius, mm |
| $r_o$ | Diaphragm radius, mm |
| $t$ | Time, s |
| $U_x, U_y$ | Phase-averaged streamwise and transverse velocity, m/s |

$U_{cl}$     Maximum jet center-line velocity at $y/d$ = 0.075, m/s

$\overline{U}_J$     Average jet velocity during expulsion cycle, m/s

$U_\infty$     Freestream velocity, m/s

$u_*$     Friction velocity, m/s

$V_j$     Maximum jet center-line velocity assigned to Womersley solution, m/s

$W_o$     The Womersley number; $d\sqrt{\omega/4\nu}$

$y^+$     Dimensionless wall distance; $u_* y/\nu$

$x, y, z$     Streamwise, transverse and spanwise cartesian coordinates with origin

at jet slot center ($x$ = 0, $y$ = 0, $z$ = 0), mm

$\Delta t$     Computational time step, s

$\delta$     Boundary layer thickness, mm

$\varepsilon$     Turbulence kinetic energy dissipation rate, m$^2$/s$^3$

$\theta$     Actuation phase angle relative to drive signal, $^\circ$

$\theta_0$     Momentum thickness of the unactuated boundary layer, *mm*

$\rho_j$     Jet fluid density, kg/m$^3$

$\rho_\infty$     Crossflow boundary layer fluid density, kg/m$^3$

$\nu$     Fluid kinematic viscosity, m$^2$/s

$\phi$     Actuation phase angle relative to mass flow rate within the neck at $y/d$

= -5, $^\circ$

$\omega$     Actuation frequency in radians; $2\pi f$

**REFERENCES**

[1]     Feero, M. A., Goodfellow, S. D., Lavoie, P., and Sullivan, P. E., 2015, "Flow

        Reattachment Using Synthetic Jet Actuation on a Low-Reynolds-Number Airfoil,"

        AIAA J., **53**(7), pp. 2005–2014.

[2]     Zhang, S., and Zhong, S., 2011, "Turbulent Flow Separation Control over a Two-

        Dimensional Ramp Using Synthetic Jets," AIAA J., **49**(12), pp. 2637–2649.

[3]     Yen, J., and Ahmed, N. A., 2012, "Parametric Study of Dynamic Stall Flow Field

        with Synthetic Jet Actuation," J. Fluids Eng. Trans. ASME, **134**(7).

[4]     Tadjfar, M., and Kamari, D., 2020, "Optimization of Flow Control Parameters over

        SD7003 Airfoil with Synthetic Jet Actuator," J. Fluids Eng. Trans. ASME, **142**(2).

[5]     Zhang, B., Liu, H., Li, Y., Liu, H., and Dong, J., 2021, "Experimental Study of Coaxial

        Jets Mixing Enhancement Using Synthetic Jets," Appl. Sci., **11**(2), pp. 1–13.

[6]     He, W., Luo, Z. bing, Deng, X., Peng, C., Liu, Q., Gao, T. xiang, Cheng, P., Zhou, Y.,

        and Peng, W. qiang, 2023, "Numerical Study on the Atomization Mechanism and

        Energy Characteristics of Synthetic Jet/Dual Synthetic Jets," Appl. Energy, **346**.

[7]     Kobayashi, R., Nishibe, K., Watabe, Y., Sato, K., and Yokota, K., 2018, "Vector

        Control of Synthetic Jets Using an Asymmetric Slot," J. Fluids Eng. Trans. ASME,

        **140**(5).

[8]     Kang, Y., Luo, Z. bing, Deng, X., Cheng, P., Peng, C., He, W., and Xia, Z. xun, 2023,

        "Numerical Study of a Liquid Cooling Device Based on Dual Synthetic Jets

        Actuator," Appl. Therm. Eng., **219**.

[9]     Gil, P., 2023, "Flow and Heat Transfer Characteristics of Single and Multiple

Synthetic Jets Impingement Cooling," Int. J. Heat Mass Transf., **201**, p. 123590.

[10]  Deng, X., Dong, Z., Liu, Q., Peng, C., He, W., and Luo, Z., 2022, "Dual Synthetic Jets Actuator and Its Applications—Part III: Impingement Flow Field and Cooling Characteristics of Vectoring Dual Synthetic Jets," Actuators, **11**(12), p. 376.

[11]  Singh, P. K., Renganathan, M., Yadav, H., Sahu, S. K., Upadhyay, P. K., and Agrawal, A., 2022, "An Experimental Investigation of the Flow-Field and Thermal Characteristics of Synthetic Jet Impingement with Different Waveforms," Int. J. Heat Mass Transf., **187**, p. 122534.

[12]  Azzawi, I. D. J., Jaworski, A. J., and Mao, X., 2021, "An Overview of Synthetic Jet under Different Clamping and Amplitude Modulation Techniques," J. Fluids Eng. Trans. ASME, **143**(3).

[13]  Ziadé, P., Feero, M. A., and Sullivan, P. E., 2018, "A Numerical Study on the Influence of Cavity Shape on Synthetic Jet Performance," Int. J. Heat Fluid Flow, **74**, pp. 187–197.

[14]  Montazer, E., Mirzaei, M., Salami, E., Ward, T. A., Romli, F. I., and Kazi, S. N., 2016, "Optimization of a Synthetic Jet Actuator for Flow Control around an Airfoil," IOP Conf. Ser. Mater. Sci. Eng., **152**(1), p. 012023.

[15]  Feero, M. A., Lavoie, P., and Sullivan, P. E., 2015, "Influence of Cavity Shape on Synthetic Jet Performance," Sensors Actuators, A Phys., **223**, pp. 1–10.

[16]  Jabbal, M., Wu, J., and Zhong, S., 2006, "The Performance of Round Synthetic Jets in Quiescent Flow," Aeronaut. J., **110**(1108), pp. 385–393.

[17]  Albright, S. O., and Solovitz, S. A., 2016, "Examination of a Variable-Diameter

Synthetic Jet," J. Fluids Eng. Trans. ASME, **138**(12).

[18]    Wang, L., and Feng, L. H., 2020, "The Interactions of Rectangular Synthetic Jets with a Laminar Cross-Flow," J. Fluid Mech.

[19]    Jankee, G. K., and Ganapathisubramani, B., 2021, "Scalings for Rectangular Synthetic Jet Trajectory in a Turbulent Boundary Layer," J. Fluid Mech., **915**.

[20]    Wen, X., and Tang, H., 2017, "Dye Visualization of In-Line Twin Synthetic Jets in Crossflows - A Parametric Study," J. Fluids Eng. Trans. ASME, **139**(9).

[21]    Salunkhe, P., Wu, Y., and Tang, H., 2020, "Aerodynamic Performance Improvement of a Wing Model Using an Array of Slotted Synthetic Jets," J. Fluids Eng. Trans. ASME, **142**(10).

[22]    Jabbal, M., and Zhong, S., 2010, "Particle Image Velocimetry Measurements of the Interaction of Synthetic Jets with a Zero-Pressure Gradient Laminar Boundary Layer," Phys. Fluids, **22**(6), pp. 1–17.

[23]    Ho, H. H., Essel, E. E., and Sullivan, P. E., 2022, "The Interactions of a Circular Synthetic Jet with a Turbulent Crossflow," Phys. Fluids, **34**(7), p. 75108.

[24]    Chapin, V. G., and Benard, E., 2015, "Active Control of a Stalled Airfoil through Steady or Unsteady Actuation Jets," J. Fluids Eng. Trans. ASME, **137**(9).

[25]    Matiz-Chicacausa, A., Molano, S., and Lopez, O. D., 2023, "Flow Control with Synthetic Jets on a Wind Turbine Airfoil," AIAA Aviat. Forum.

[26]    Tousi, N. M., Bergadà, J. M., and Mellibovsky, F., 2022, "Large Eddy Simulation of Optimal Synthetic Jet Actuation on a SD7003 Airfoil in Post-Stall Conditions," Aerosp. Sci. Technol., **127**, p. 107679.

[27]   Guoqiang, L., and Shihe, Y., 2020, "Large Eddy Simulation of Dynamic Stall Flow

       Control for Wind Turbine Airfoil Using Plasma Actuator," Energy, **212**.

[28]   Pasa, J., Panda, S., and Arumuru, V., 2023, "Focusing of Jet from Synthetic Jet

       Array Using Non-Linear Phase Delay," Phys. Fluids, **35**(5).

[29]   Panda, S., Gohil, T. B., and Arumuru, V., 2022, "Influence of Mass Flux Ratio on

       the Evolution of Coaxial Synthetic Jet," Phys. Fluids, **34**(9).

[30]   Palumbo, A., Semeraro, O., Robinet, J. C., and De Luca, L., 2022, "Boundary Layer

       Transition Induced by Low-Speed Synthetic Jets," Phys. Fluids, **34**(12).

[31]   Wang, Y. Z., Mei, Y. F., Aubry, N., Chen, Z., Wu, P., and Wu, W. T., 2022, "Deep

       Reinforcement Learning Based Synthetic Jet Control on Disturbed Flow over

       Airfoil," Phys. Fluids, **34**(3).

[32]   Bai, H., Wang, F., Zhang, S., Zhang, W., and Lin, Y., 2023, "Square Cylinder Flow

       Controlled by a Synthetic Jet at One Leading Edge," Phys. Fluids, **35**(3).

[33]   Kral, L. D., Donovan, J. F., Cain, A. B., and Cary, A. W., 1997, "Numerical

       Simulation of Synthetic Jet Actuators," 4th Shear Flow Control Conf.

[34]   Belanger, R., Zingg, D. W., and Lavoie, P., 2020, "Vortex Structure of a Synthetic

       Jet Issuing into a Turbulent Boundary Layer from a Finite-Span Rectangular

       Orifice," AIAA Scitech 2020 Forum, **1 PartF**.

[35]   Ciobaca, V., Rudnik, R., Haucke, F., and Nitsche, W., 2013, "Active Flow Control on

       a High-Lift Airfoil: URANS Simulations and Comparison with Time-Accurate

       Measurements," 31st AIAA Appl. Aerodyn. Conf.

[36]   Asgari, E., and Tadjfar, M., 2022, "Role of Phase-Difference between Two

Adjacent Rectangular Synthetic Jet Actuators in Active Control of Flow over a Rounded Ramp," Phys. Fluids, **34**(2).

[37]  Kotapati, R. B., Mittal, R., and Cattafesta, L. N., 2007, "Numerical Study of a Transitional Synthetic Jet in Quiescent External Flow," J. Fluid Mech., **581**, pp. 287–321.

[38]  Yao, C., Chen, F. J., and Neuhart, D., 2006, "Synthetic Jet Flowfield Database for Computational Fluid Dynamics Validation," AIAA J., **44**(12), pp. 3153–3157.

[39]  Coskun, U. C., Cadirci, S., and Gunes, H., 2021, "Numerical Investigation of Active Flow Control on Laminar Forced Convection over a Backward Facing Step Surrounded by Multiple Jets," J. Appl. Fluid Mech., **14**(2), pp. 447–458.

[40]  Wang, Y. Z., Mei, Y. F., Aubry, N., Chen, Z., Wu, P., and Wu, W. T., 2022, "Deep Reinforcement Learning Based Synthetic Jet Control on Disturbed Flow over Airfoil," Phys. Fluids, **34**(3).

[41]  Raju, R., Aram, E., Mittal, R., and Cattafesta, L., 2009, "Simple Models of Zero-Net Mass-Flux Jets for Flow Control Simulations," Int. J. Flow Control, **1**(3), pp. 179–197.

[42]  Launder, B. E., and Sharma, B. I., 1974, "Application of the Energy-Dissipation Model of Turbulence to the Calculation of Flow near a Spinning Disc," Lett. Heat Mass Transf., **1**(2), pp. 131–137.

[43]  Jasak, H., 2009, "OpenFOAM: Open Source CFD in Research and Industry," Int. J. Nav. Archit. Ocean Eng., **1**(2), pp. 89–94.

[44]  Feero, M. A., Lavoie, P., and Sullivan, P. E., 2017, "Influence of Synthetic Jet

Location on Active Control of an Airfoil at Low Reynolds Number," Exp. Fluids, **58**(8), p. 99.

[45]  Celik, I. B., Ghia, U., Roache, P. J., Freitas, C. J., Coleman, H., and Raad, P. E., 2008, "Procedure for Estimation and Reporting of Uncertainty Due to Discretization in CFD Applications," J. Fluids Eng. Trans. ASME, **130**(7), pp. 0780011–0780014.

[46]  Hunt, J. C. R., Wray, A. A., and Moin, P., 1988, "Eddies, Streams, and Convergence Zones in Turbulent Flows," Cent. Turbul. Res. Proc. Summer Progr., (1970), pp. 193–208.

[47]  New, T. H., Lim, T. T., and Luo, S. C., 2006, "Effects of Jet Velocity Profiles on a Round Jet in Cross-Flow," Exp. Fluids, **40**(6), pp. 859–875.

[48]  Zhong, S., Garcillan, L., and Wood, N. J., 2005, "Dye Visualisation of Inclined and Skewed Synthetic Jets in a Cross Flow," Aeronaut. J., **109**(1093), pp. 147–155.

[49]  Sabatino, D. R., and Maharjan, R., 2015, "Characterizing the Formation and Regeneration of Hairpin Vortices in a Laminar Boundary Layer," Phys. Fluids, **27**(12).

[50]  Xu, K., Lavoie, P., and Sullivan, P., 2023, "Flow Reattachment on a NACA 0025 Airfoil Using an Array of Microblowers," AIAA J., **61**(6), pp. 2476–2485.

[51]  Greenblatt, D., Paschal, K. B., Yao, C. S., and Harris, J., 2006, "Experimental Investigation of Separation Control Part 2: Zero Mass-Flux Oscillatory Blowing," AIAA J., **44**(12), pp. 2831–2845.

**Figure Captions List**

Fig. 1    Schematic drawing of a synthetic jet actuator in a crossflow during ingestion cycle

Fig. 2    Schematic drawing of different SJA implementations. Red line for neck-only and jet-slot-only case indicate the boundary which the jet profile is applied.

Fig. 3    Mesh configuration and boundary conditions for the whole SJA case in a crossflow. (a) Symmetry plane view with coordination axis (only part of the overall length is displayed) and (b) zoomed view of the SJA.

Fig. 4    Comparison of centre-line jet velocity sampled near the jet exit ($y/d$ = 0.075) over one actuation cycle in quiescent. Note: neck-only case is shifted linearly 10degrees to match peak phase.

Fig. 5    SJA radial jet velocity profile during (a) peak expulsion and (b) peak ingestion phases, sampled at $y/d$ = 0.075. Experimental data from Feero et al.[15].

Fig. 6    Instantaneous Q-Criterion contours, colored with normalized streamwise velocity. (Threshold: $Q^* = Qd^2/U_\infty^2 = 0.01$) VR: Vortex ring; HR: Horseshoe vortex; TVP: Trailing vortex pair. NTV: near-wall tertiary vortex.

Fig. 7    Phase-averaged streamline plots over streamwise velocity contour

Fig. 8    Time-averaged (a) normalised streamwise velocity and (b) turbulence kinetic energy comparison along the symmetry plane.

Fig. 9        Time-averaged streamwise velocity comparison of $x$-$z$ plane sampled at $y/d$ = 0.5

Fig. 10       $x$-$z$ plane time-averaged streamwise velocity deviation comparison between the whole SJA and jet-slot-only case, sampled at $y/d$ = 0.5.

Fig. 11       Time-averaged streamwise velocity comparison along the spanwise direction at $y/d$ = 2.5.

Fig. 12       Time-averaged skin friction coefficient contour comparison of the bottom wall

Fig. 13       Instantaneous (a) streamwise and (b) transverse velocity measured at jet exit center with increasing neck height.

Fig. 14       Convergence of phase-averaged (a) Streamwise (b) Transverse velocity jet profile with increasing neck height.

Fig. 15       Comparison of streamwise and transverse velocity at the bottom of the neck, normalised by respective maximum centre-line velocity applied for the BC.

Fig. 16       Comparison of (a) turbulence kinetic energy and (b) turbulence dissipation rate, at bottom of the neck during peak ingestion phase.

**Table Caption List**

Table 1     Summary of grid properties for mesh sensitivity study of synthetic jet in

            crossflow

Table 2     Volume ratio comparison of the neck only perimetric study cases

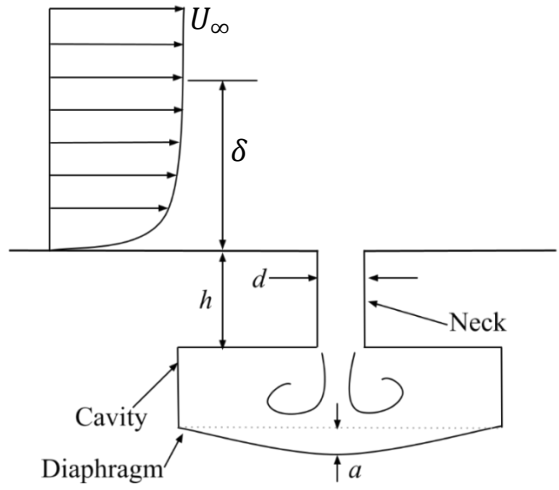

**Fig. 1 Schematic drawing of a synthetic jet actuator in a crossflow during ingestion cycle.**

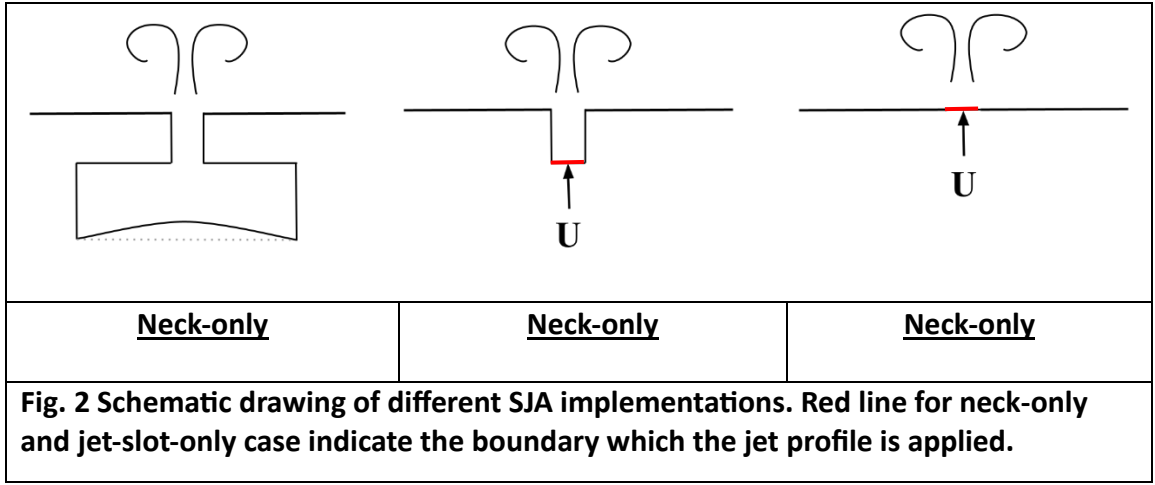

| Neck-only | Neck-only | Neck-only |

**Fig. 2 Schematic drawing of different SJA implementations. Red line for neck-only and jet-slot-only case indicate the boundary which the jet profile is applied.**

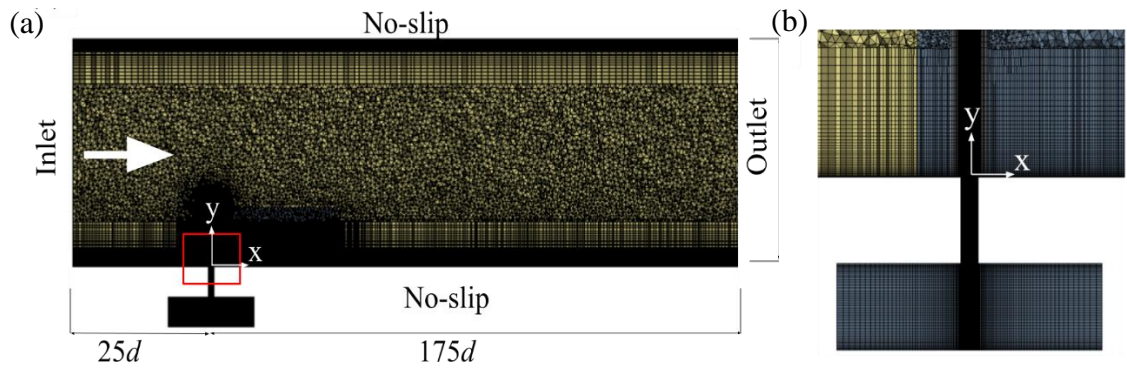

**Fig. 3**. Mesh configuration and boundary conditions for the whole SJA case in a crossflow. (a) Symmetry plane view with coordination axis (only part of the overall length is displayed) and (b) zoomed view of the SJA.

**TABLE 1 Summary of grid properties for mesh sensitivity study of synthetic jet in crossflow.**

| Grid | Total Cell Count | SJA cell count | $U_{cl}$ $(m/s)$ | $U_{cl}$ Uncertainty (%) | $\delta^*_{x/d=10}$ | $\delta^*_{x/d=10}$ Uncertainty (%) |
|---|---|---|---|---|---|---|
| **Whole-I** | 1.97M | 59k | 9.664 | - | 1.665 | |
| **Whole-II** | 4.46M | 130k | 10.168 | 3.1 | 1.684 | 1.6 |
| **Whole-III** | 10.89M | 320k | 10.345 | 1.0 | 1.675 | <1 |
| **Neck-I** | 1.92M | 10k | 9.787 | - | 1.660 | - |
| **Neck-II** | 4.35M | 22k | 10.237 | 1.4 | 1.682 | 1.2 |
| **Neck-III** | 10.61M | 50k | 10.248 | <1 | 1.672 | <1 |

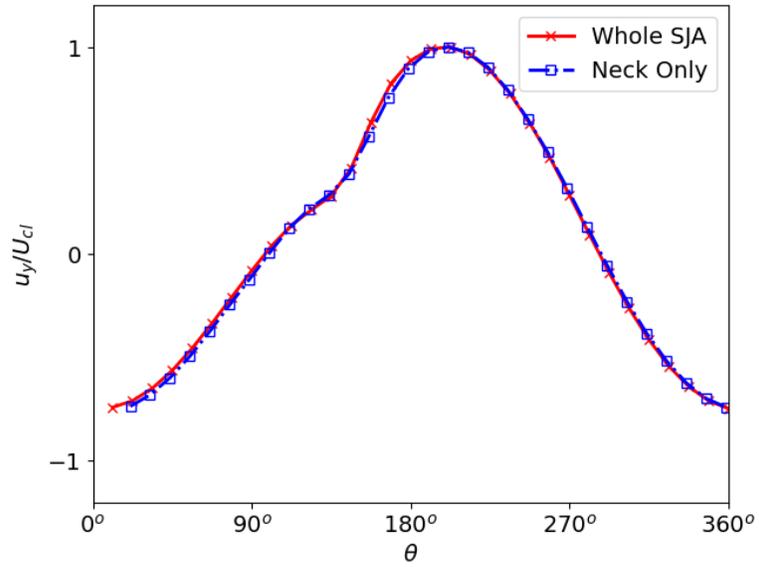

**Fig. 4 Comparison of center-line jet velocity sampled near the jet exit (*y/d* = 0.075) over one actuation cycle in quiescent. Note: neck-only case is shifted linearly 10degrees to match peak phase.**

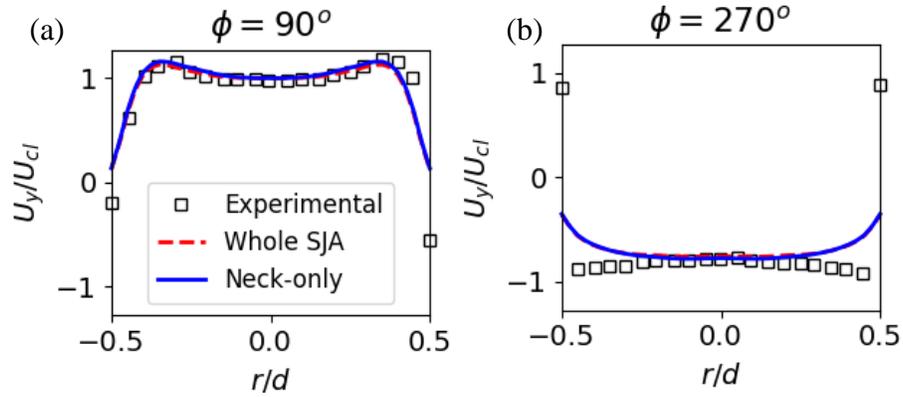

**Fig. 5 SJA radial jet velocity profile during (a) peak expulsion and (b) peak ingestion phases, sampled at *y/d* = 0.075. Experimental data from Feero et al.[15].**

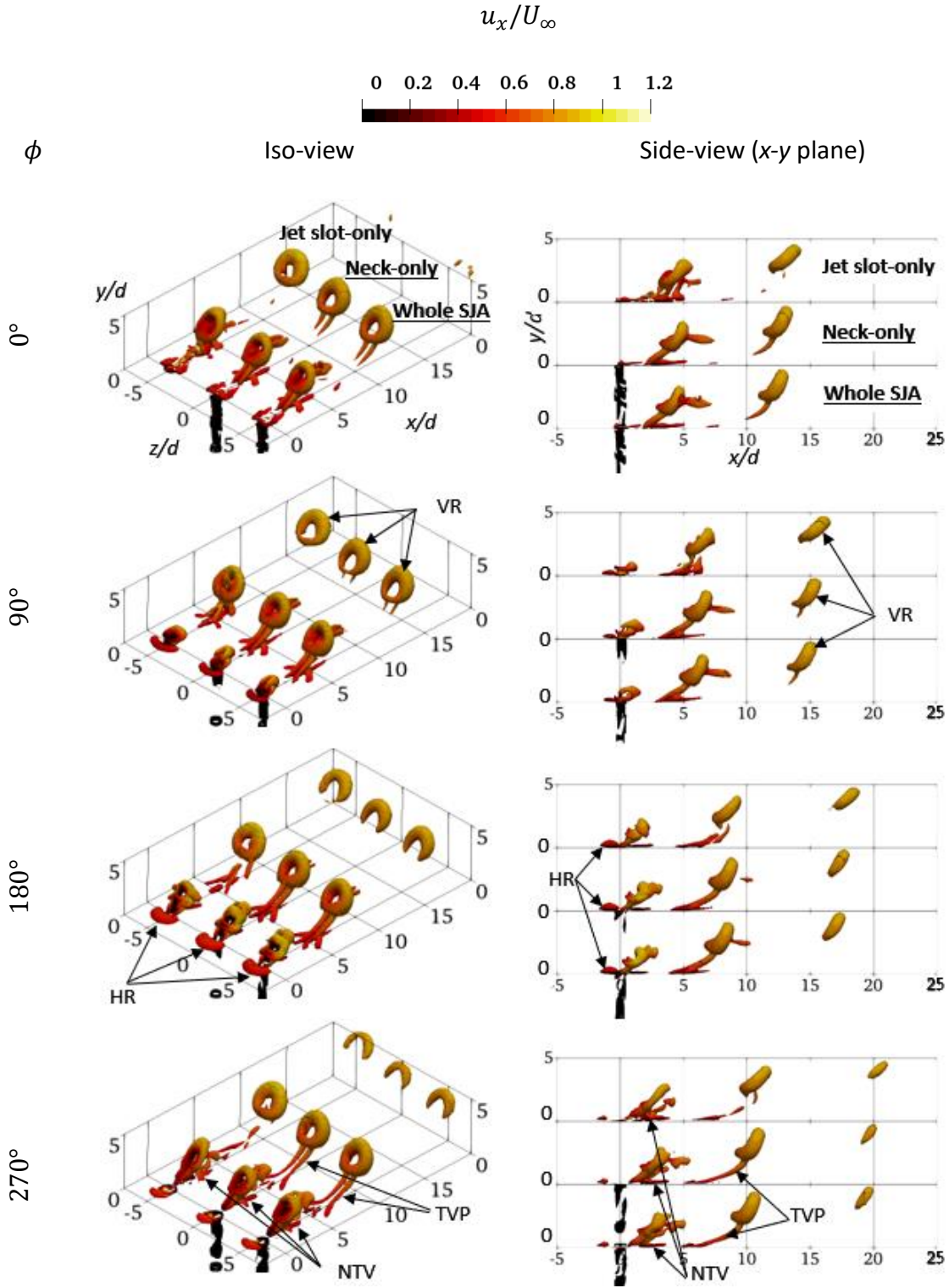

**Fig. 6 Instantaneous Q-Criterion contours, coloured with normalised streamwise velocity. (Threshold: $Q^* = Qd^2/U_\infty^2 = 0.01$) VR: Vortex ring; HR: Horseshoe vortex; TVP: Trailing vortex pair. NTV: near-wall tertiary vortex.**

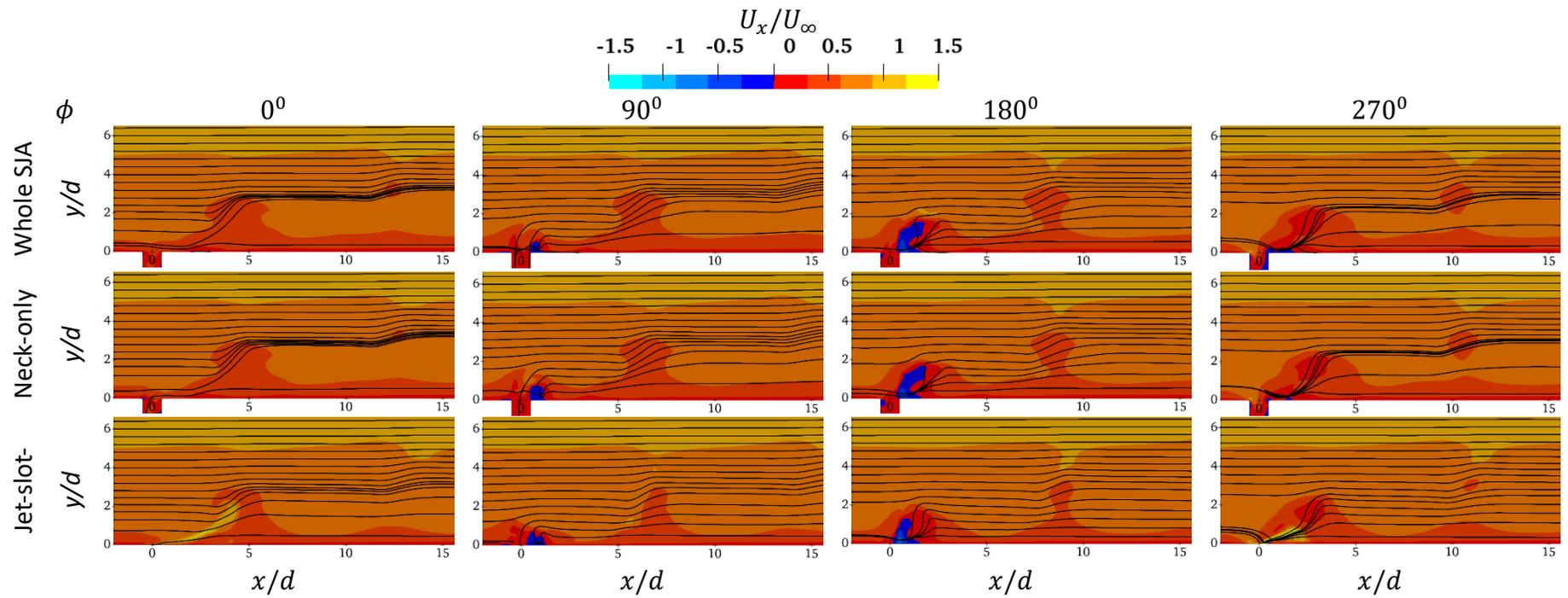

Fig. 7 Phase-averaged streamline plots over streamwise velocity contour.

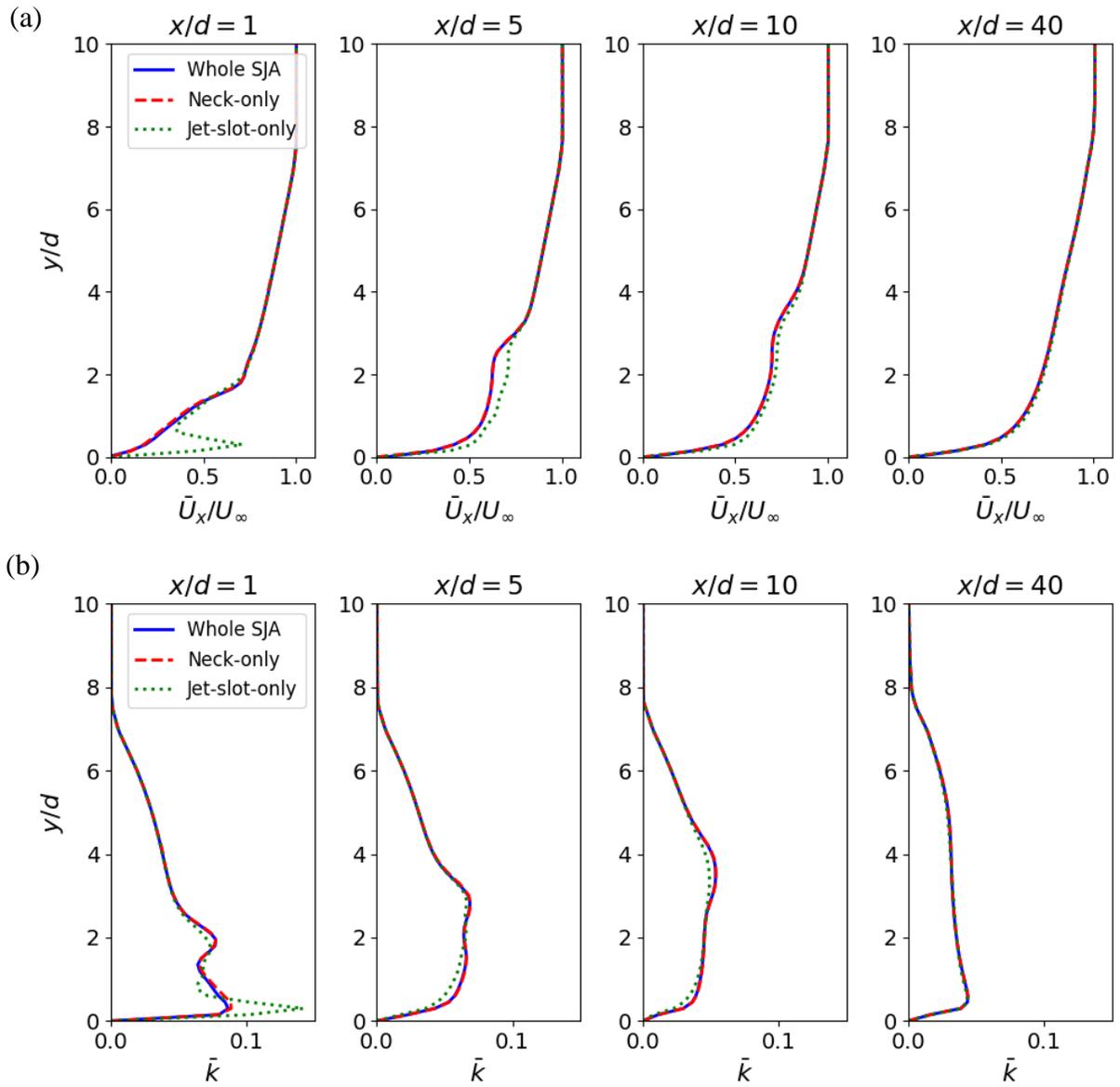

**Fig. 8** Time-averaged (a) normalised streamwise velocity and (b) turbulence kinetic energy comparison along the symmetry plane.

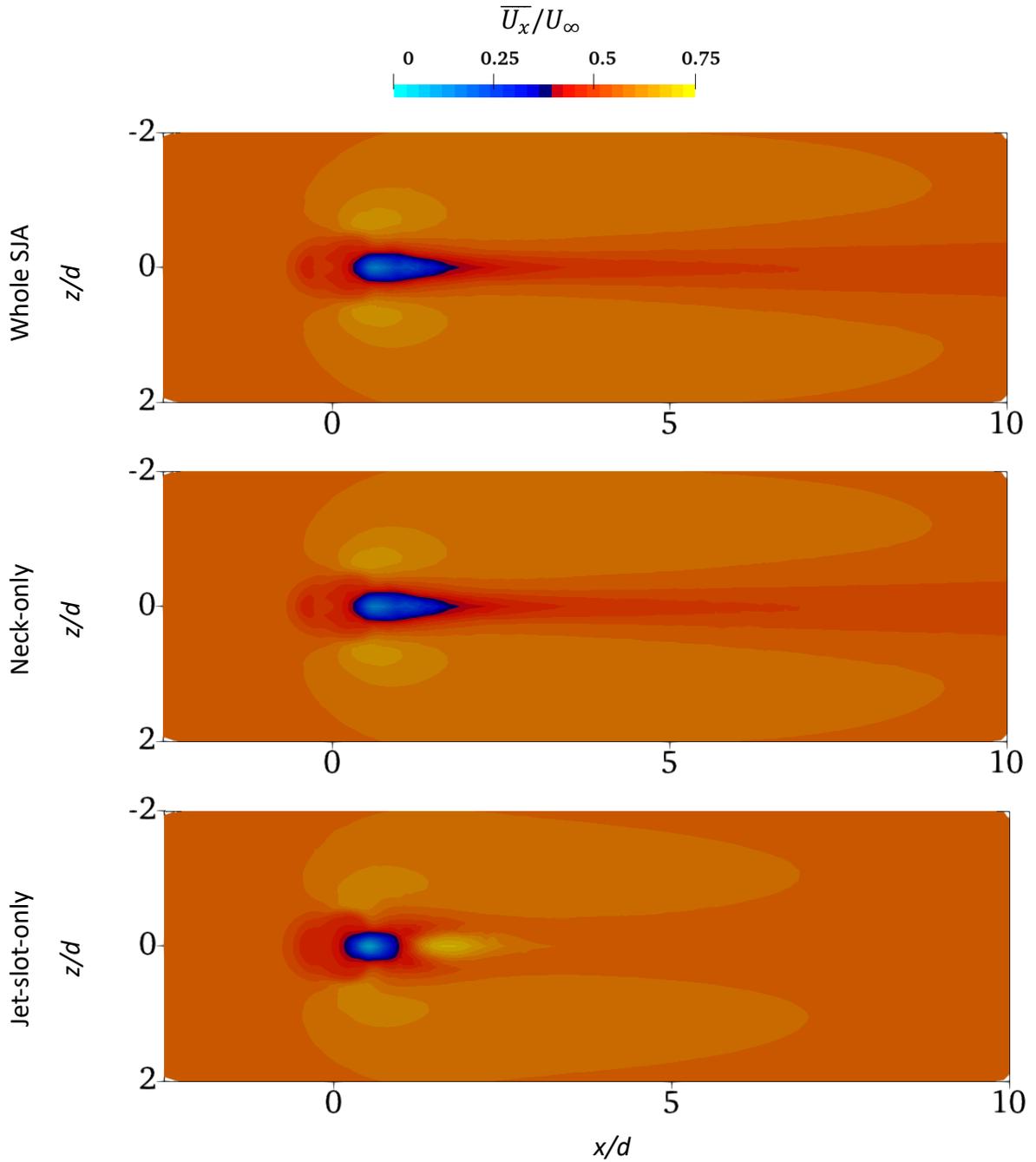

**Fig. 9 Time-averaged streamwise velocity comparison of *x-z* plane sampled at *y/d* = 0.5.**

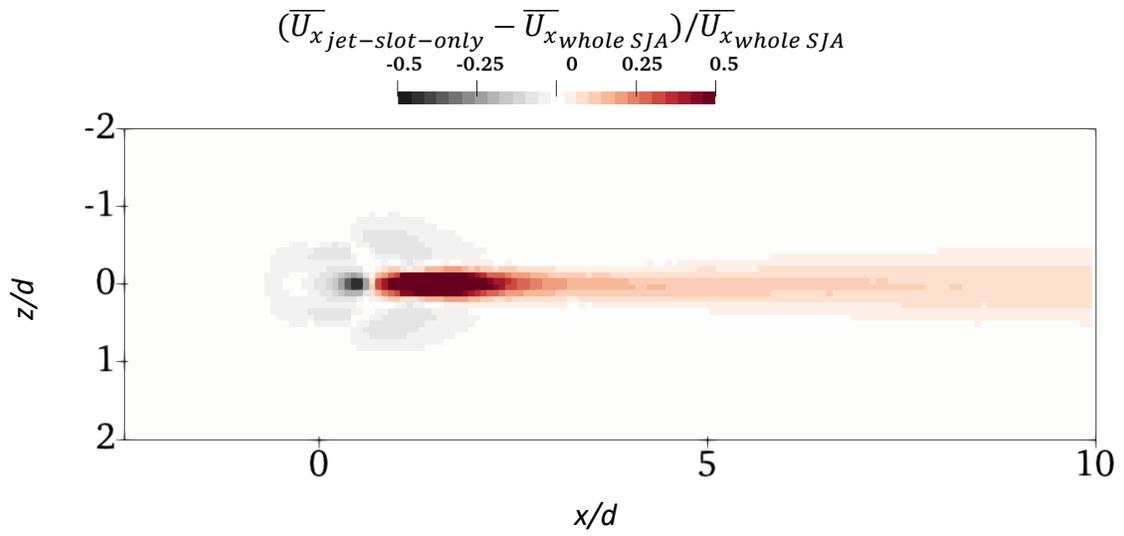

**Fig. 10** *x-z* plane time-averaged streamwise velocity deviation comparison between the whole SJA and jet-slot-only case, sampled at *y/d* = 0.5.

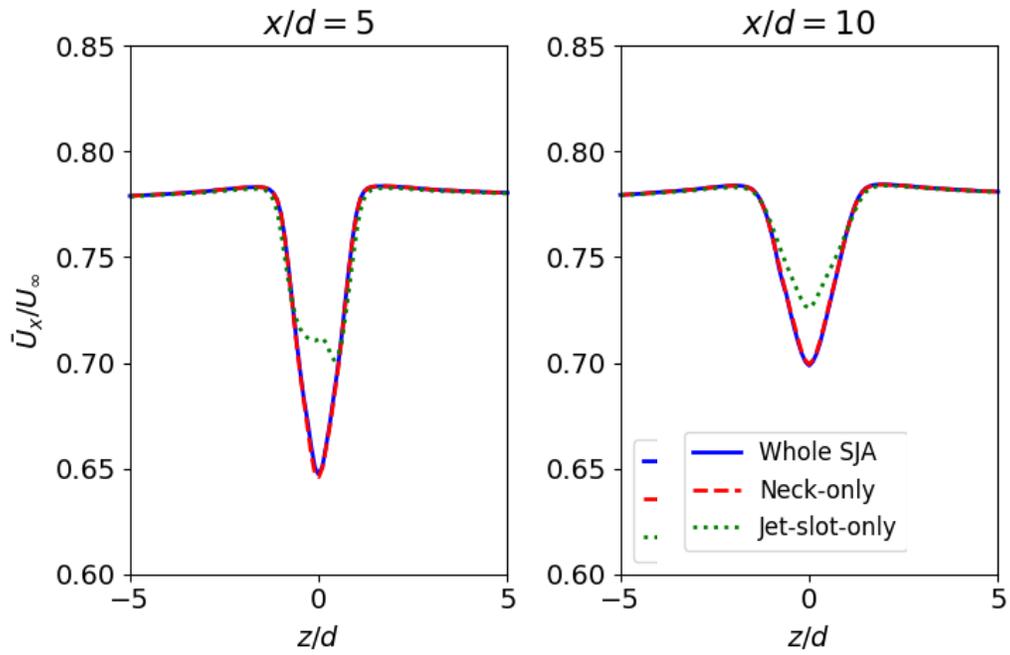

**Fig. 11 Time-averaged streamwise velocity comparison along the spanwise direction at *y/d* = 2.5.**

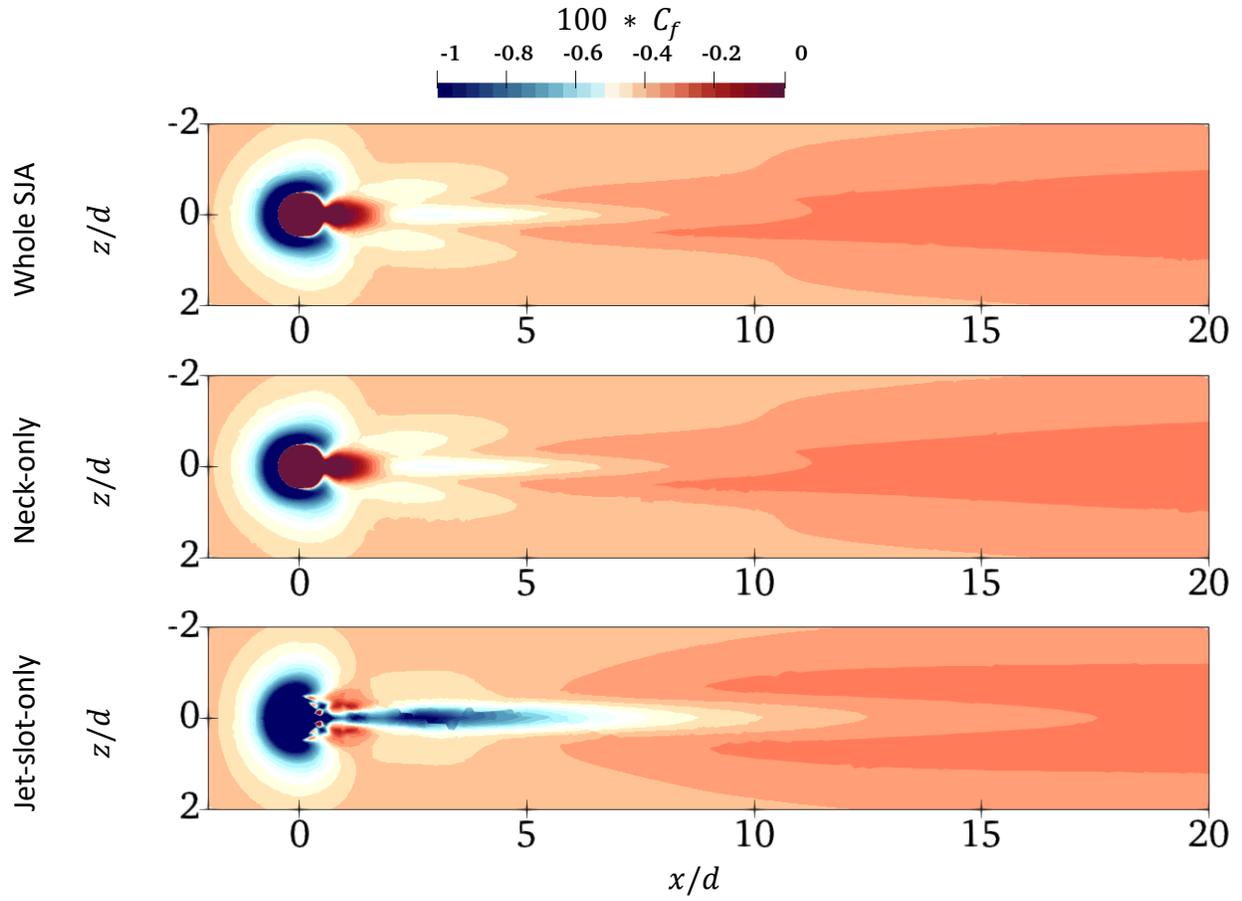

**Fig. 12 Time-averaged skin friction coefficient contour comparison of the bottom wall.**

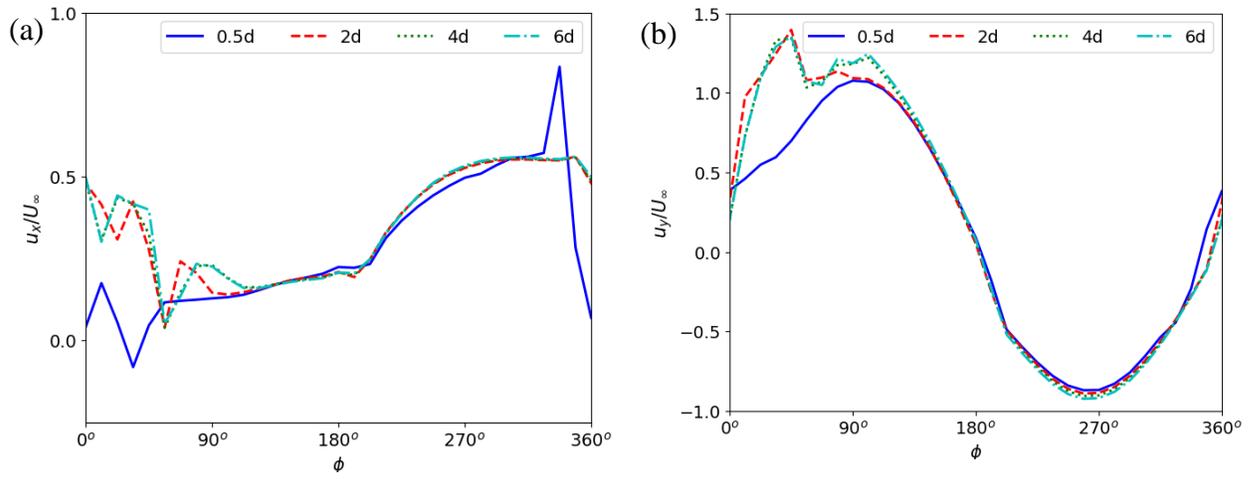

**Fig. 13 Instantaneous (a) streamwise and (b) transverse velocity measured at jet exit center with increasing neck height.**

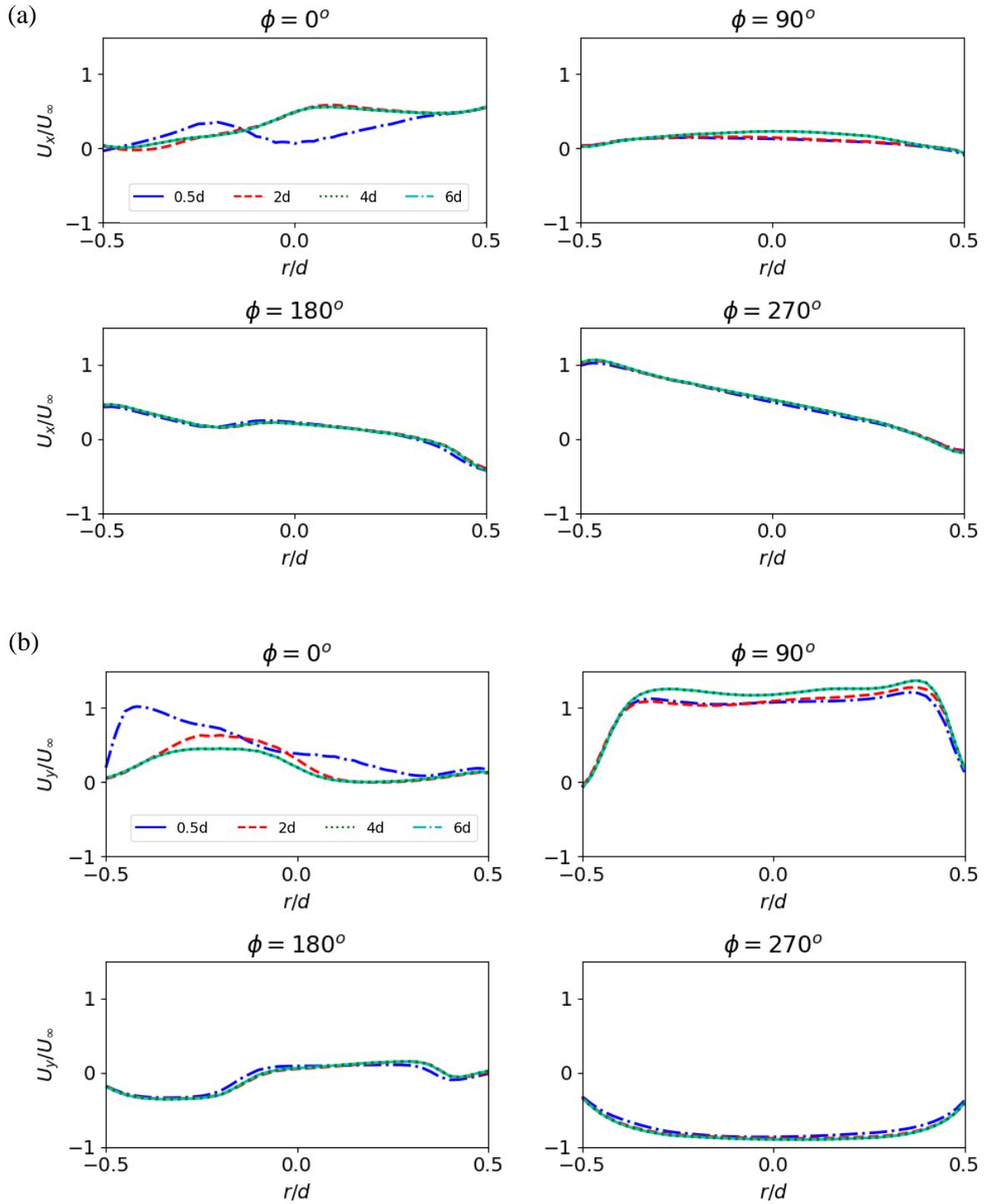

**Fig. 14 Convergence of phase-averaged (a) Streamwise (b) Transverse velocity jet profile with increasing neck height.**

**TABLE 2 Volume ratio comparison of the neck only perimetric study cases**

| Neck height modeled ($h/d$) | Volume Ratio |
|---|---|
| 0.5 | 7.58 |
| 2 | 1.89 |
| 4 | 0.95 |
| 5 (Whole SJA) | 0.76 |
| 6 | 0.63 |

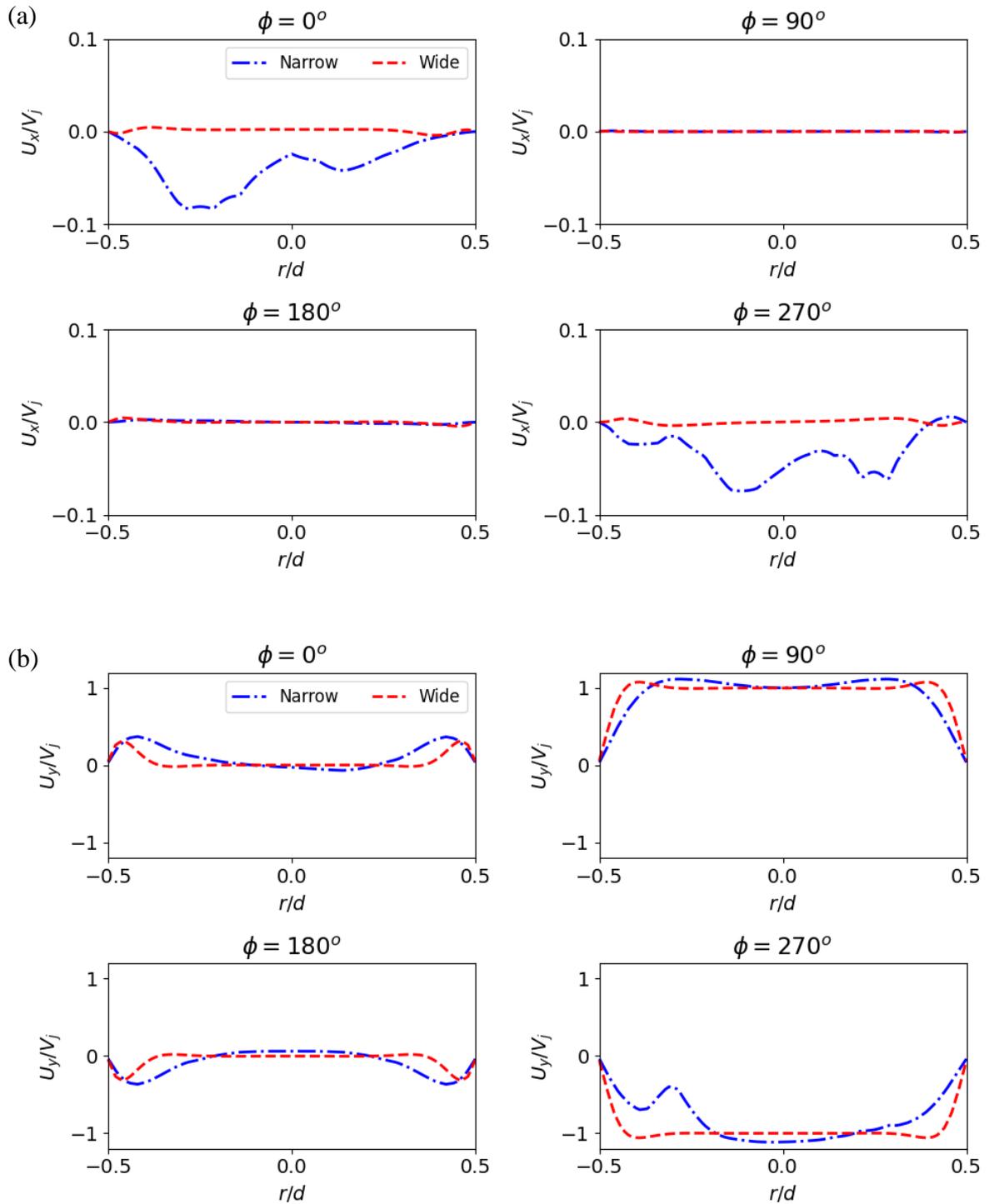

**Fig. 15 Comparison of streamwise and transverse velocity at the bottom of the neck, normalised by respective maximum center-line velocity applied for the BC.**

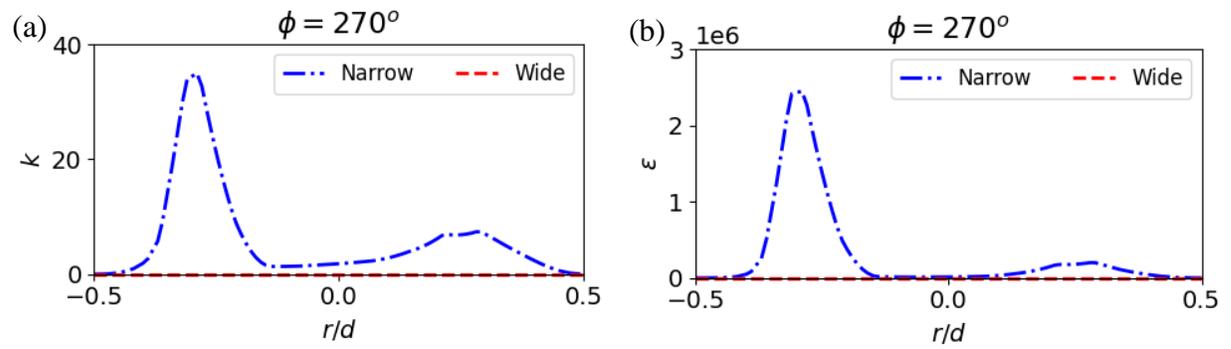

**Fig. 16 Comparison of (a) turbulence kinetic energy and (b) turbulence dissipation rate, at bottom of the neck during peak ingestion phase.**